# Correlations of non-affine displacements in metallic glasses through the yield transition

Richard Jana[1,2], Lars Pastewka[1,2,3,4,*]

[1] Department of Microsystems Engineering, University of Freiburg, Georges-Köhler-Allee 103, 79110 Freiburg, Germany

[2] Institute for Applied Materials, Karlsruhe Institute of Technology, Straße am Forum 4, 76131 Karlsruhe, Germany

[3] Freiburg Materials Research Center, University of Freiburg, Stefan-Meier-Straße 21, 79104 Freiburg, Germany

[4] Cluster of Excellence *liv*MatS @ FIT – Freiburg Center for Interactive Materials and Bioinspired Technologies, University of Freiburg, Georges-Köhler-Allee 105, 79110 Freiburg, Germany

* Corresponding author: lars.pastewka@imtek.uni-freiburg.de

**Abstract**

We study correlations of non-affine displacement during simple shear deformation of Cu-Zr bulk metallic glasses in molecular dynamics calculations. In the elastic regime, our calculations show exponential correlation with a decay length that we interpret as the size of a shear transformation zone in the elastic regime. This correlation length becomes system-size dependent beyond the yield transition as our calculation develops a shear band, indicative of a diverging length scale. We interpret these observations in the context of a recent proposition of yield as a first-order phase transition.

## 1. Introduction

Bulk metallic glasses (BMG), multi-component metals that are kinetically arrested into an amorphous structure, have been suggested for wide range of applications, including as structural materials[1,2]. For practical applications a big problem is their tendency to form shear bands, planar regions in that localize most of the plastic deformation at relatively low strain. These shear bands are the primary mechanism by which metallic glasses fail. Numerous ideas to address this problem have been suggested, such as deliberately introducing pores[3] or creating "nanoglasses" that have an internal microstructure[4].

The deformation of BMGs is described by the theory of shear transformations or shear transformation zones (STZs)[5–7], localized rearrangements of small regions of atoms. The size of these zones has been estimated to range from a few[8] to many tens of atoms[9,10]. Knowledge of the size of the zones could help to fundamentally understand this class of materials on an atomic level and be used in mesoscale simulations that incorporate STZs[11–13]. The size of STZs has been linked to the Poisson ratio[14] as well as the brittle or ductile character of fracture of BMGs[15–17].

Spatial correlation functions of non-affine deformation have recently been employed to quantify the geometry of STZs. Murali et al.[18] looked at the spatial autocorrelation in the non-affine deformation field of deformed bulk metallic glasses in molecular dynamics simulations. They found an exponential decay of the autocorrelation from which they extracted a correlation length $\ell$, which they interpreted as the size of an STZ. These findings have been confirmed by similar calculation on Lennard-Glasses[19]. In a similar spirit, Chikkadi et al.[20,21] have discussed the autocorrelation of non-affine deformation in experiments of sheared colloidal glasses. In addition to the global non-affine displacement field, they characterized the local nonaffine deformation through the $D^2_{\min}$ measure of Falk & Langer[7]. Their data shows long-range correlations as manifested in a power-law behavior of the autocorrelation function global and local measures for non-affinity. In contrast to Murali et al.'s data[18], this suggests a scale-free character of the deformation. Calculations of hard-sphere mixtures carried out for the interpretation of these experiments did again yield an exponential decay of the correlation function[21,22]. Varnik et al.[23] argued that this is because of limitation in system size; larger calculation, albeit carried for a 2D soft disk model rather than in 3D, indeed showed power-law correlations. Earlier calculations of the correlation of the vorticity field during deformation of a 2D Lennard-Jones solid showed similar power-law correlations[24,25].

We here revisit the question of exponential vs. power-law correlations and provide new data on how they evolve through the yield transition. Our molecular dynamics calculations of the deformation of BMGs show the emergence of correlations in the non-affine part of the deformation field of calculation larger than those previously reported. This allows us to extract the correlation function up to distance ~75 times the nearest neighbor distance for the largest systems studied here, similar to previous 2D calculations that showed power-law correlation[23]. While we do find exponential and not power-law correlations, the length-scale $\ell$ associated with the exponential becomes a function of system size after shear-band nucleation, indicating a divergent length at the nucleation of the band.

## 2. Methods

### 2.1. Molecular dynamics simulations

We conducted all simulations using Molecular Dynamics (MD) and the second generation of the interatomic Cu-Zr potential by Mendelev et al.[26]. Amorphous sample systems were first obtained by melting and equilibrating systems of $Cu_{50}Zr_{50}$ (or other stoichiometries, see below) at 2500 K for 100 ps, followed by a linear quench to 750 K at a rate of 6 K ps$^{-1}$. This temperature is slightly above the glass transition temperature $T_g \approx 600$ K, as obtained from the jump in heat capacity when cooling the system through $T_g$ at the same rate. We then aged the system for 1 ns before quenching it to 0 K at a rate of 6 K ps$^{-1}$. We used a Berendsen barostat[27] with a relaxation time constant of 10 ps to keep the hydrostatic pressure in the simulation cell at zero and Langevin thermostat[28] with a relaxation time constant of 1 ps to control temperature during quench and equilibration.

To prepare simulations carried out at different temperatures, the amorphous systems were then equilibrated at zero pressure for 200 ps at different temperatures between 0 and 300 K. The cell was subsequently deformed using simple shear deformation at constant volume at an applied shear rate of $\dot{\varepsilon} = 10^8$ s$^{-1}$ up to a maximum of 35% strain. To control temperature, we again used a Langevin thermostat but only thermalize the Cartesian direction normal to the plane of shear to eliminate any drag with respect to some reference velocity field. The bulk of our simulations comprises a cubic cell with an edge length of $L \approx 206$ Å and 500,000 atoms. The potential influence of finite-size effects was studied using two additional system sizes: A cubic system with twice the edge length and eight times the number of atoms and another cubic system with half the edge length and 1/8 the number of atoms. If not mentioned otherwise, results are reported for the $L = 206$ Å system.

## 2.2. Local strain measure and correlation

To quantify heterogeneous flow of the system, we need measures that identify local deformation events. Falk & Langer[7] introduced a method to determine the local deformation of an atomic system within spheres of radius $r_{cut}$. The idea is to map for each atom $i$ its atomic neighborhood at time $t-\Delta t$ to the neighborhood at time $t$ using an affine deformation with deformation gradient $\underline{F_i}$, and then find $\underline{F_i}$ that minimizes the residual error. The final residual error,

$$D_{\min,i}^2 = \min_{\underline{F_i}} \left\{ \frac{1}{N} \sum_k^N [\boldsymbol{r}_{ik}(t) - \underline{F_i} \cdot \boldsymbol{r}_{ik}(t-\Delta t)]^2 \theta(r_{\text{cut}} - r_{ik}) \right\}, \tag{1}$$

is a measure for the non-affine component of local deformation. Here $\Theta(r)$ is the Heaviside step function. Shear transformation zones and shear bands can be identified by looking for regions with high values of $D_{\min,i}^2$. Note that for our calculations carried out at constant applied shear rate $\dot{\varepsilon}$, we specify the reference frame by its distance in applied strain $\Delta\varepsilon$ rather than $\Delta t \equiv \Delta\varepsilon/\dot{\varepsilon}$.

To quantify the geometry of these deformation events, we use spatial auto-correlation maps. The auto-correlation map of some field $Q(\vec{r})$ is defined as

$$\mathcal{A}[Q](\vec{r}) = V \int d^3 r' \, Q(\vec{r}')Q^*(\vec{r}-\vec{r}') = \frac{V}{N_p^2} \sum_i \sum_j Q_i Q_j^* \, \delta(\vec{r} - \vec{r}_{ij}) \tag{2}$$

where the last identity is the expression obtained for $N_p$ point particles for which $Q(\vec{r}) = \sum_i Q_i \delta(\vec{r} - \vec{r}_i)$ where $Q_i$ is the value of quantity $Q$ on atom $i$. Note that for any quantity $Q$ this autocorrelation map obeys the following sum rules,

$$\mathcal{A}[Q](\vec{r} \to 0) = \langle Q_i^2 \rangle \quad \text{and} \quad \mathcal{A}[Q](\vec{r} \to \infty) = \langle Q_i \rangle^2 \tag{3}$$

where the average $\langle Q_i \rangle = \sum_i Q_i / N_p$.

The radial average of this auto-correlation map gives a function $\mathcal{A}[Q](r)$, depending only on the distance and not the direction between the two atoms. The auto-correlation function of unity is the radial distribution (or pair correlation) function,

$$g_2(r) = \mathcal{A}[1](r). \tag{4}$$

By virtue of Eq. (3) it is normalized such that $g_2(r) \to 1$ as $r \to \infty$. We are specifically interested in the correlations of $D_{min}^2$,

$$C(\vec{r}) = \frac{\mathcal{A}[D_{min}^2](\vec{r}) - \langle D_{min,i}^2\rangle^2}{\langle(D_{min,i}^2)^2\rangle - \langle D_{min,i}^2\rangle^2}. \tag{5}$$

Note that Eq. (5) is normalized such that, because of Eq. (3), $C(\vec{r} \to 0) = 1$ and $C(\vec{r} \to \infty) = 0$. We compute $\mathcal{A}[Q](\vec{r})$ at short distances by directly evaluating Eq. (2) and at long distances using a fast Fourier transform to speed up the convolution in Eq. (2), allowing us to efficiently compute the correlation function up to half the size of our systems. We have implemented this algorithm in matscipy[29] and Ovito[30].

## 3. Results

Figure 1a shows a snapshot of the quenched system before shear. The radial distribution function $g_2(r)$ (Fig. 2a) is indicative of a disordered structure with broad nearest and second-nearest neighbor peaks and non-zero probability for finding a neighbor between them. The first neighbor peak is located at $r_{NN} = 2.8$Å and indicated by a vertical dashed line. The value of the non-affine displacement $D_{min,i}^2$ depends on the cutoff distance $r_{cut}$ for identifying neighbors of an atom and on the distance $\Delta\varepsilon$ between current and reference configuration in the time domain. In the following, we will show results obtained for $r_{cut}$ being an integer multiple of the nearest-neighbor distance $r_{NN}$ as given by the position of the first peak in $g_2(r)$. These distances are indicated by the vertical dashed lines in Fig. 2a.

After equilibrating, we deformed our metallic glass under simple shear. Figures 1b and c show exemplary snapshots of these calculations. The mean square displacement in the $z$-direction, perpendicular to the plane where shear is applied (Fig. 2b), shows that the system was supercooled and atoms broke out of their cages at a strain between 0.1% and 1% (see e.g. Refs. 31,32). The shear stress $\sigma_{xy}$ in the plane of shear (Fig. 3a) initially rose linearly with the strain $\varepsilon$ applied in the $xy$-plane. At around $\varepsilon \approx 10\%$ the system yielded and the stress $\sigma_{xy}$ dropped from a peak value to a plateau region where the $\sigma_{xy}$ remained constant up to an applied strain of $\varepsilon = 35\%$, the maximum strain applied in our calculations. Our five calculations at 0 K, 50 K, 100 K, 200 K and 300 K show that the system softened as temperature increased; from a yield stress of around 1.7 GPa in the athermal limit to 1.2 GPa at 100 K.

Figures 3b-d show a map of $D^2_{\min}$ during deformation, here computed for a cutoff $r_{\text{cut}} = 3\ r_{\text{NN}}$ and a reference frame at an applied strain $\Delta\varepsilon = 1\%$, about the cage-breaking strain, before the frame shown in the figure. At small strain $\varepsilon$ where $\sigma_{xy}(\varepsilon)$ is linear, we find localized events (Fig. 3b). After yield, these localized events coalesce to shear-bands, first vertical (Fig. 3c, see also Ref. 19) but later horizontal (Fig. 3d), developing a clear anisotropic structure. Note that such vertical shear-bands occurred only in some of our calculations. At small strain, both shear-band directions are equivalent and the nucleation direction is random. Symmetry breaking at larger strain forces the shear band back into the direction of shear.

To statistically quantify this (random) structure we computed the $D^2_{\min}$ auto-correlation maps, $C(\vec{r})$. Figure 4a and b show a slice $C(x, y, z = 0) \equiv C_0(x, y)$. Before yield (Fig. 4a), $C_0(x, y)$ shows a rotationally symmetric structure with a visible ring at the nearest-neighbor distance $r_{\text{NN}}$. After yield (Fig. 4b), $C_0(x, y)$ develops a clear anisotropic structure with a band of large correlation parallel to the $x$-axis. Fig. 4c shows radial averages $C(r)$ of the data of Figs. 4a and b. There are oscillations at small distances that turn into an exponential decay at around 10 Å. Oscillations at small distances are due to the structure of the amorphous solid. We therefore normalize the autocorrelation function and define

$$\bar{C}(r) = C(r)/g_2(r) \tag{6}$$

to remove variations in $C(r)$ due to variations in local atomic density. Figure 4d shows that these oscillations are eliminated in $\bar{C}(r)$ for $r > 5$Å.

In the following, we characterize the exponential decay,

$$\bar{C}(r) = A\, exp(-r/\ell), \tag{7}$$

by fitting the correlation length $\ell$ in Eq. (7) over a select section of the correlation function. We distinguish between the behavior at short distances 5 Å $< r <$ 15 Å, which is within the range of cutoff radii $r_{cut}$ we used for the computation of $D^2_{min}$. We denote the corresponding correlation length by $\ell_{short}$. The behavior at long distances is fitted to the region 20 Å $< r <$ 30 Å and we denote the correlation length by $\ell_{long}$.

Note that the computation of $\bar{C}(r)$ involves the cutoff radius $r_{cut}$ as a parameter. $r_{cut}$ determines the local atomic neighborhood within which $D^2_{min}$ is calculated. To test whether the length scale $\ell$ depends on this length, we parametrically vary $r_{cut}$ between $r_{cut} = 2\, r_{NN} = 5.6$Å and $r_{cut} = 5\, r_{NN} = 14$ Å. The resulting correlation functions at 7% and 12% applied strain are shown in Figs. 5a and b, respectively. The radius $r_{cut}$ varies by a factor of 2.5 while the individual correlation functions move systematically upwards. As a consequence, the extracted value $\ell_{short}$ depends systematically on $r_{cut}$. Indeed, we can collapse all $\ell_{short}$ values on a single curve when normalizing by $r_{cut}$, $\ell_{short}/r_{cut}$ (Fig. 5c). The behavior of $\ell_{long}$ is different. Its value (Fig. 5d) is independent of $r_{cut}$ used in the computation of $D^2_{min}$ and the data does not collapse when normalized accordingly. The evolution of $\ell_\text{short}$ and $\ell_\text{long}$ with applied strain $\varepsilon$ clearly show the point where the samples yield (cf. also Fig. 3a). At around 12.5% strain, $\ell_\text{long}$ increases dramatically. During further deformation it fluctuates around a value consistently a factor of 3 higher than before yield.

The computation of $\bar{C}(r)$ furthermore depends on how the reference frame for the computation of $D^2_\text{min}$ is chosen. Here, we report results obtained for references frames at constant distance in applied strain, $\Delta\varepsilon$. All results reported above were obtained for $\Delta\varepsilon = 1\%$. Fig. 6 demonstrates how the correlation function and $\ell$ vary as a function of this parameter. Before yield, the correlation function does not depend on $\Delta\varepsilon$ and drops exponentially over two decades as a function of distance. Fig. 6a shows this behavior for $\Delta\varepsilon = 1\%$, 0.1% and 0.01% which is above, at and below the cage-breaking strain (Fig. 2b). The behavior changes at yield (Fig. 6b), where the initial exponential drop starts to depend on $\Delta\varepsilon$. Fig. 6c shows the influence on the extracted value of $\ell_\text{long}$. $\ell_\text{long}$ decreases with decreasing $\Delta\varepsilon$ and saturates at $\ell_\text{long} \approx 15$ Å in the flow region for the lowest $\Delta\varepsilon = 0.01\%$.

To clarify the role of $\Delta\varepsilon$ on the calculation of the correlation length $\ell$, we further test the influence on system size on the correlation functions. Fig. 7a shows that before yield (applied strain $\varepsilon = 7\%$), $\bar{C}(r)$ is independent of system size but that a clear size dependence develops when the material flows (Fig. 7c, $\varepsilon = 20\%$). Plotting $\ell_\text{long}$ versus applied strain $\varepsilon$ shows that the before yield $\ell_\text{long}$ is independent of size but after yield it depends on system size (Fig. 7b). Normalizing distance $r$ or correlation length $\ell_\text{long}$ by system size collapses all data in the region where the amorphous solid flows (Fig. 7c and d).

Finally, we test the dependence of the correlation function on temperature and composition. Fig. 8a shows the temperature dependence of $\ell_{\text{long}}$. Data in the temperature range from 50K to 300K, below the glass transition temperature of $T_g \approx 600K$ of our metallic glass, is superimposed for small strain. It appears that at large strain the higher temperatures lead to a smaller $\ell_{\text{long}}$ but out present data is too noisy to make a firm conclusion. Fig 8b shows $\ell_{\text{long}}(\varepsilon)$ for different compositions. Again the data collapses in the elastic regime and there appears to be a slight variation with composition after the sample has yielded.

**Discussion**

The correlation length $\ell$ characterizing the exponential decay of the spatial-autocorrelation functions $\bar{C}(r)$ of $D_{\text{min}}^2$ have in the past been interpreted as giving the size of the STZ[18]. Our results clearly show that the decay of $\bar{C}(r)$ with distance $r$ is exponential in molecular dynamics calculations of BMGs, confirming other results obtained for EAM[18], Lennard-Jones[19] and hard-sphere glasses[21,22]. However, there are two regions of exponential decay with different correlation lengths. At short distance $r < r_{\text{cut}}$, the characteristic length $\ell_{\text{short}}$ is strongly affected by the choice of $r_{\text{cut}}$ within which the nonaffine part of the local deformation field is computed. Our results indicate $\ell_{\text{short}} \propto r_{\text{cut}}$ such that $\ell_{\text{short}}$ does not characterize any intrinsic material scale. The initial decay crosses over to a second exponential at distances $r > r_{\text{cut}}$ with a characteristic length $\ell_{\text{long}}$ that does not depend on the specific choice of $r_{\text{cut}}$ and reference frame and is a characteristic of the material under investigation. For the CuZr glasses investigated here we find $\ell_{\text{long}} \sim 5 - 10$ Å. This is on the order of the values reported for FeP in Ref.[18] ($\ell = 8.5$ Å) but smaller than the values for MgAl ($\ell = 11.1$ Å) and CuZr ($\ell = 15.0$ Å) reported there at an applied strain of $\varepsilon = 4\%$ for simulations carried out with an earlier version of the EAM potential used here[33].

Additionally, Ref.[18] used the initial configuration at $\varepsilon = 0$ as reference and looked at correlations of global nonaffine displacements rather than $D_{min}^2$. Recent work using a Lennard-Jones model for CuZr reports $\ell = 5$ Å [19]. While this appears to indicate that the actual value of the correlation length is highly model-dependent and may also depend on the preparation of the glass, we find that the values extracted from our calculations are robust to variations of temperature and stoichiometry.

The situation before yield is characterized by individual regions of large $D_{min}^2$ (Fig. 1b) that are typically attributed to individual STZs. Therefore, $\bar{C}(r)$ measures the autocorrelation of the deformation field of an *individual* STZ. Since the overall density of STZs is low, the strain offset $\Delta\varepsilon$ that determines over how many STZs we average does not affect the results. The situation changes dramatically after the sample has yielded ($\varepsilon > 10\%$). STZs are now localized within a shear band and it becomes difficult to identify individual STZs (Fig. 3c and d). The onset of shear-banding is then accompanied by a characteristic length $\ell_{long}$ proportional to the system size $L$ and that depends on strain offset $\Delta\varepsilon$. For small $\Delta\varepsilon$, we find values for $\ell$ comparable to the ones found in the elastic regime $\ell$ (Fig. 6c). We hypothesize, that this is because even for the flowing glass we can pick out individual STZs if we look at small enough strain increments, much smaller than the cage-breaking strain (Fig. 2b).

We note that while in the elastic regime our correlation functions look clearly exponential, our system sizes albeit large are yet too small to rule out power-law behavior during flow. Indeed, the fact that our length scale $\ell_{long}$ depends on system size is indicative of a diverging length or a cross-over to a power-law as STZ events become correlated within the shear band. This observation is consistent with a recently proposition that yield in amorphous solids can be interpreted as a first-order phase transitions[34,35], an interpretation that has a rich history for explaining shear-banding instabilities in non-Newtonian fluids[36]. Jaiswal et al.[34] identify the transition using an order parameter that measures similarity or "overlap" of atomic configuration. The atomic configuration uses overlap with the initial configuration at yield. A central observation is that their "yield" point occurs at larger strains than the overshoot in the stress-strain curve that is typically attributed to yield. This is consistent with our calculations, which show that $\ell$ rises after the stress has peaked (cf. Fig. 3a and 7b,d).

## 4. Summary & Conclusion

We studied the correlation between nonaffine displacements, as characterized by the $D_{\min}^2$ measure of Falk & Langer[7], using molecular dynamics calculations. This multipoint correlation function shows exponential behavior in the elastic regime from which we can extract a length scale $\ell$, typically attributed to the size of an STZ. We find that this length scale diverges at yield, as manifested by a size-dependent $\ell$ in during flow of the material. The diverges of $\ell$ occurs at strains larger than the peak stress that is typically attributed to the yield point. Our results support a recent proposition that yield in amorphous materials can be interpreted as a first-order phase transition[34,35].

**Acknowledgements**

We thank Suzhi Li and Jan Mees for useful discussion and the Deutsche Forschungsgemeinschaft for funding this work through grant PA 2023/2. All simulations were conducted with LAMMPS[37] on JURECA at the Jülich Supercomputing Center (grant "hfr13") and on NEMO at the University of Freiburg (DFG Grant No. INST 39/963-1 FUGG). Post-processing and visualization was carried out with ASE[38], matscipy[29] and Ovito[30].

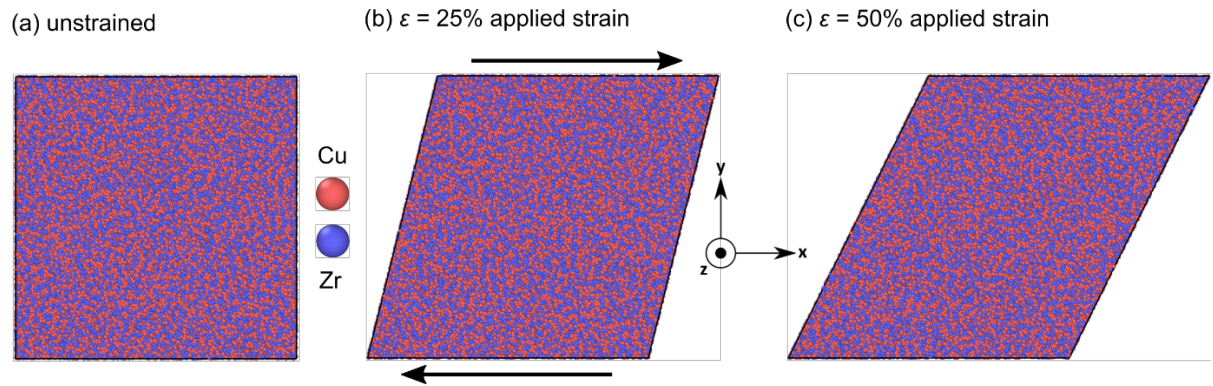

**Figure 1:** Snapshots of the system at (a) 0, (b) 25 and (c) 50% applied simple shear strain. Arrows indicate the shearing direction.

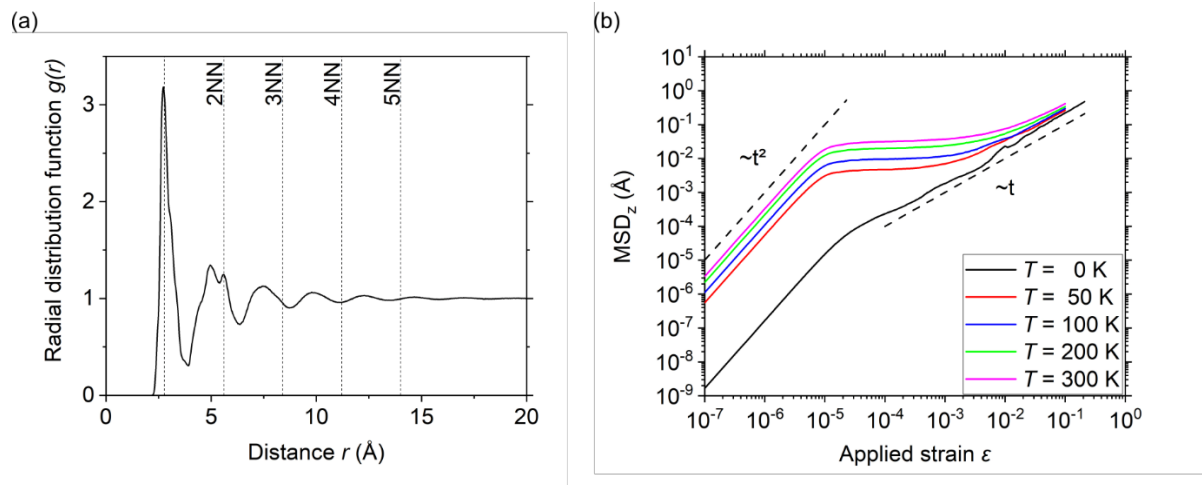

**Figure 2:** (a) Radial distribution function of the CuZr BMG at 0 K. Vertical lines represent multiples of the nearest neighbor distance used as $d_{cut}$ for the calculation of $D^2_{min}$. (b) Mean squared displacements in the z direction (perpendicular to the simple shear plane). Dashed lines show $\propto t^2$ (diffusive) and $\propto t$ (ballistic) scaling.

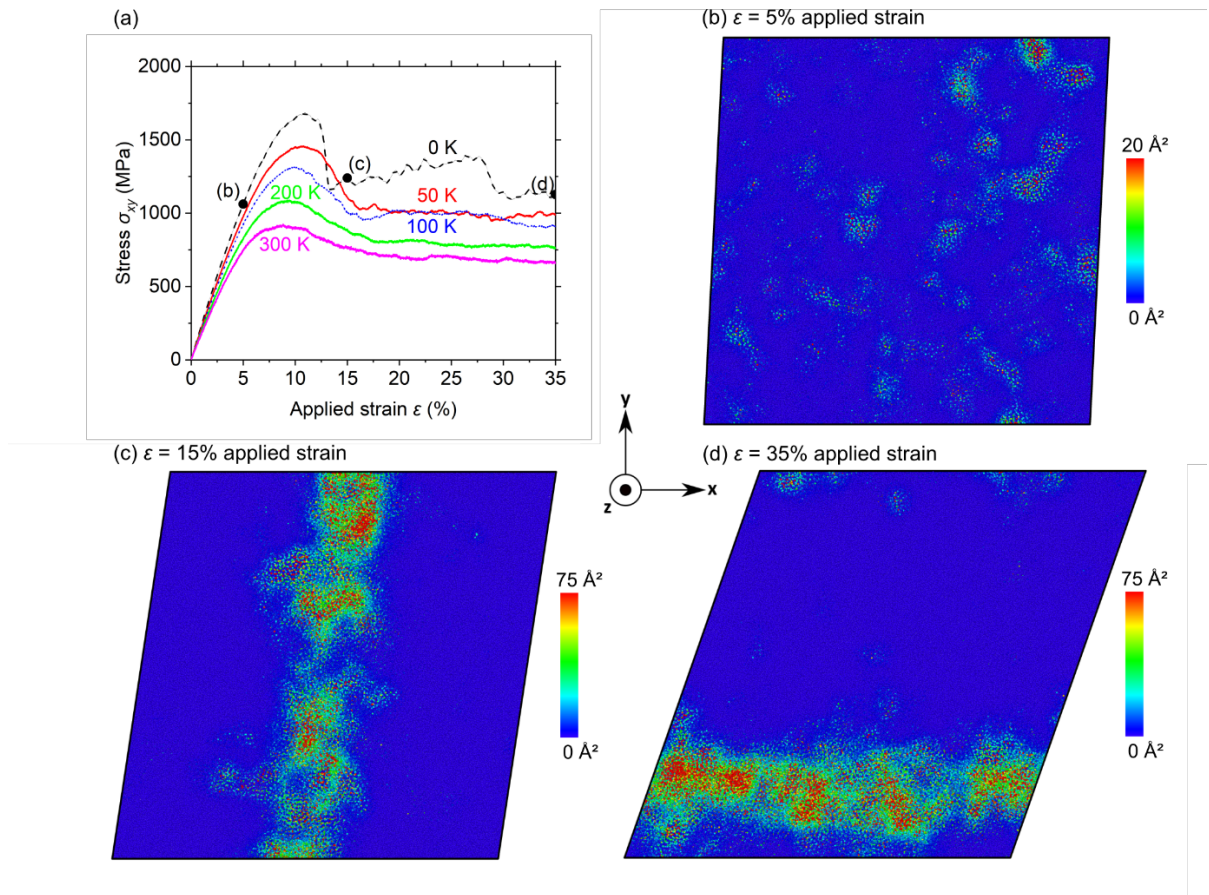

**Figure 3:** (a) Stress strain curves for CuZr at 0K, 50K, 100K, 200K and 300K. Black solid dots indicate the positions where the snapshots shown in (b)-(d) were taken. All calculations use $r_{cut} = 3\ r_{NN}$ and $\Delta\varepsilon = 1\%$. The color code corresponds to $D^2_{min}$ with high values in red and low values in blue. At low strains (b) we find individual STZs. Higher strains ((c) and (d)) develop a clear shear band.

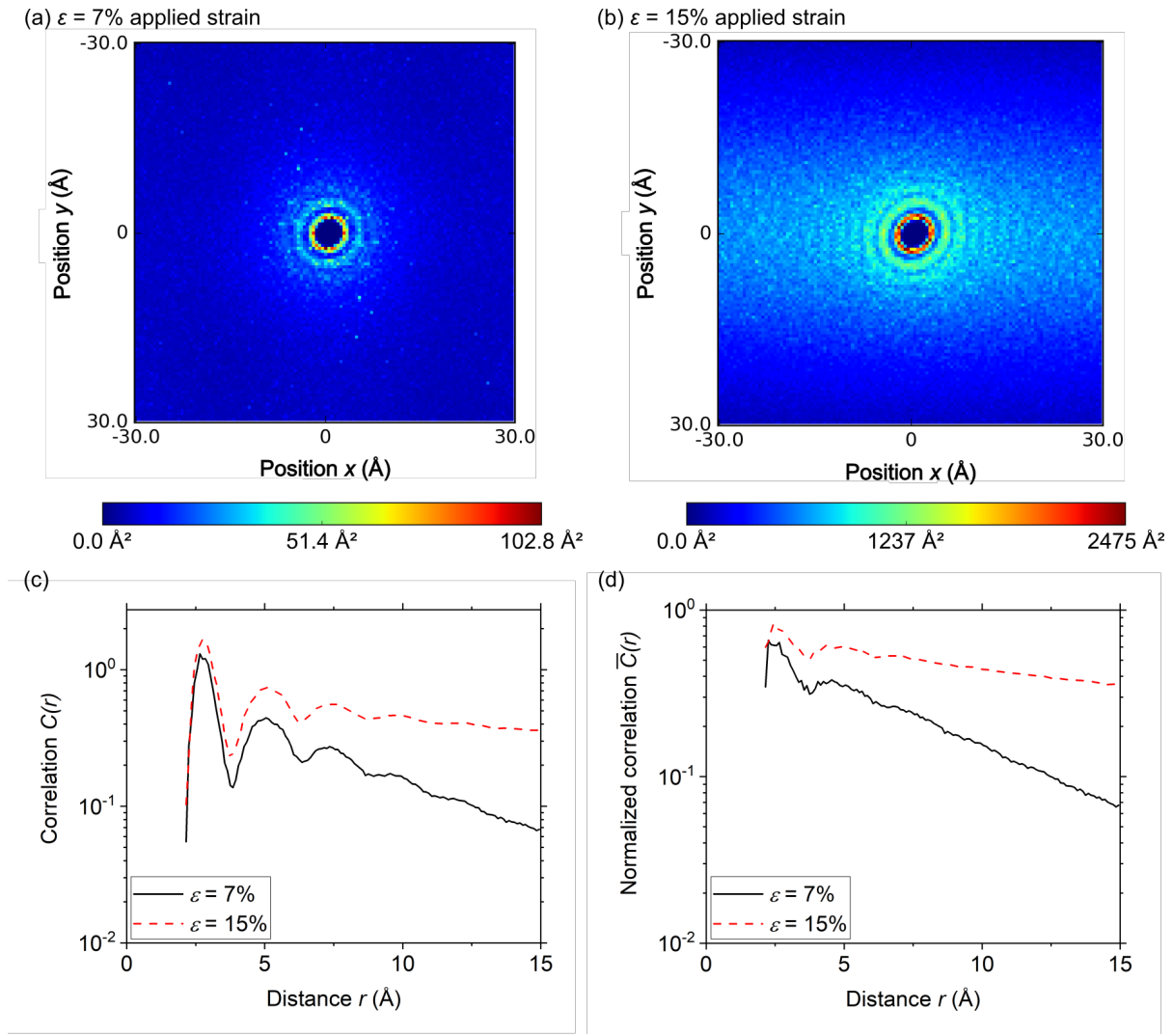

**Figure 4:** Slice through the normalized real space correlation in xy-plane at (a) 7% applied strain and (b) 15% applied strain. (c) Correlation function $C(r)$ for the two cases shown in panels (a) and (b). (d) shows the correlation function divided by the pair correlation function, $\bar{C}(r) = C(r)/g_2(r)$. All results are obtained with an offset $\Delta\varepsilon = 1\%$.

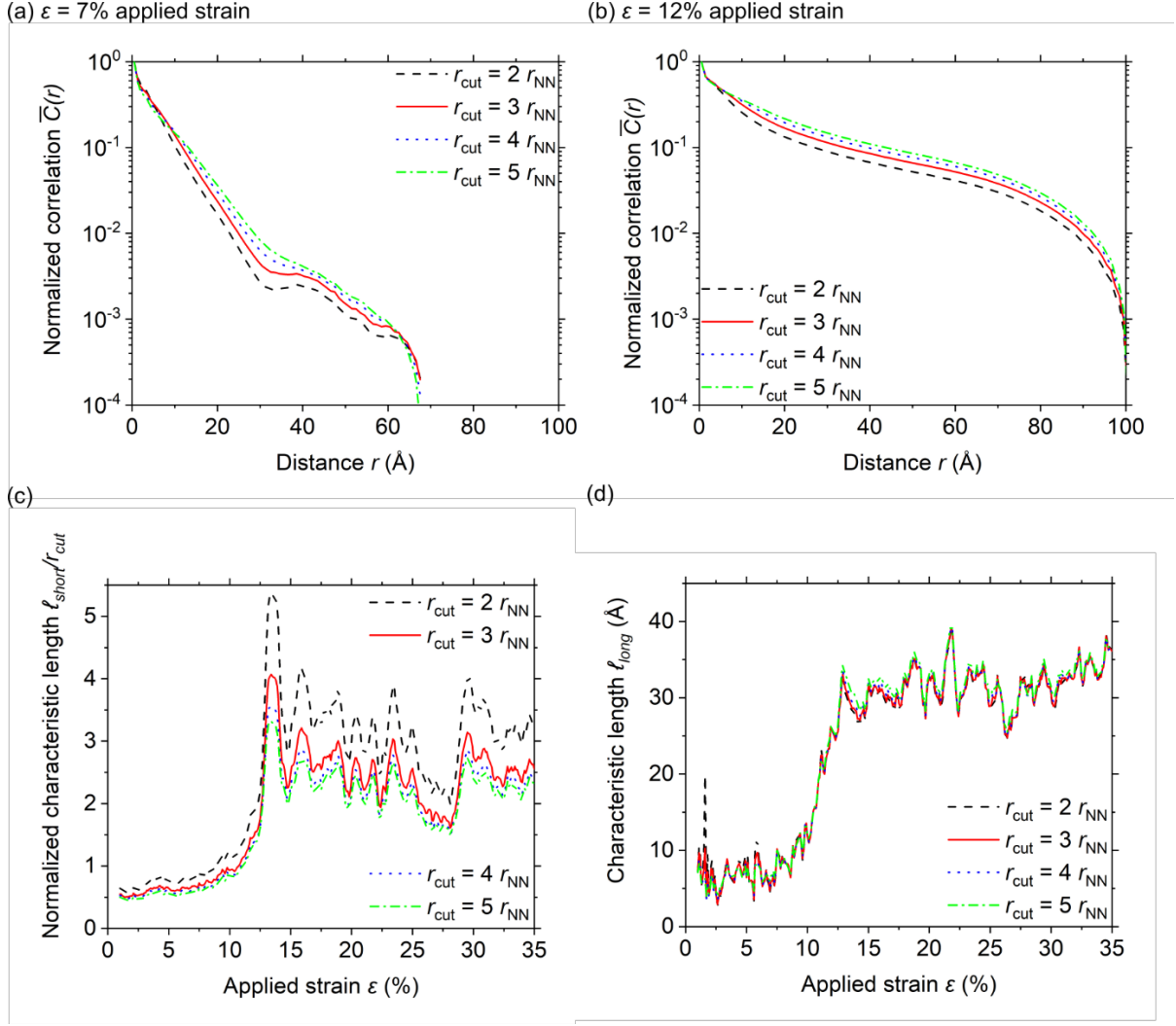

**Figure 5:** $D^2_{min}$ auto-correlation functions at (a) 7% and (b) 12% strain, using different cutoff values $r_{cut}$. (c) Characteristic length $\ell_{short}$ derived from the correlations for the different cutoffs, normalized with the cutoff $r_{cut}$ for each line. (d) Characteristic length $\ell_{long}$ derived from the correlations for the different cutoffs. All results are obtained with an offset $\Delta\varepsilon = 1\%$.

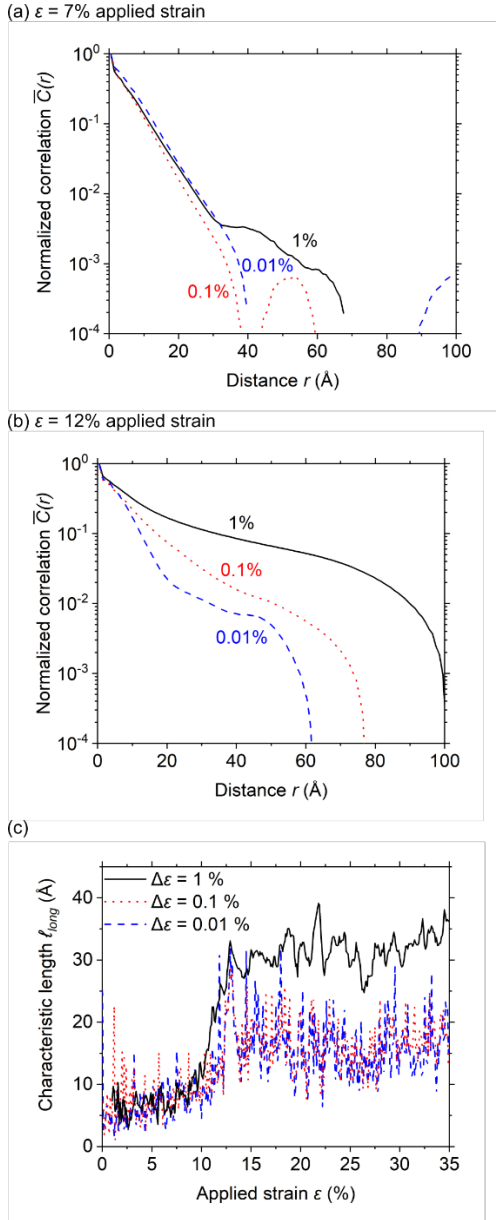

**Figure 6:** Auto-correlation functions of $D^2_{min}$ calculated over different amounts of applied strain between configurations, at 7% (a) and 12% (b) strain. (c) shows the characteristic length $\ell_{long}$ derived from the correlations for the different strain offsets. All results were obtained with $r_{cut} = 3\ r_{NN}$.

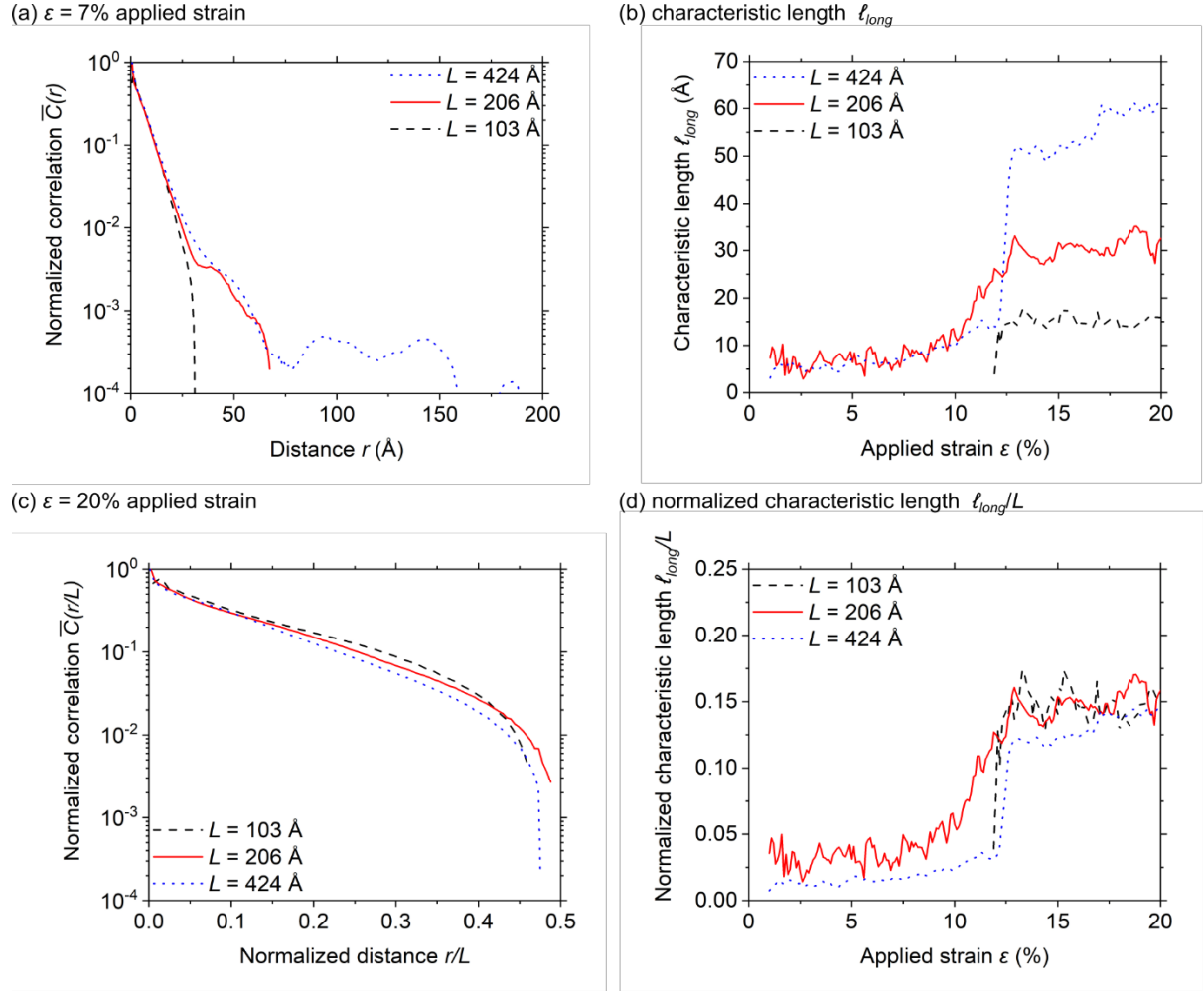

**Figure 7:** $D^2_{min}$ auto-correlation functions for systems of different sizes, at 7% (a) and 20% (c) global strain. (b) shows the characteristic length $\ell_{long}$ derived from the correlations for the systems of different size. (d) shows the same curves as (b), but normalized with the system size $L$. All results are obtained with an offset $\Delta\varepsilon = 1\%$ and $r_{cut} = 3\ r_{NN}$. $\ell_{long}$ curves for the small system with $L = 103$Å start at $\varepsilon = 11.9\%$ because the data could not be fit to exponential over the range from 20 Å to 30 Å used to extract $\ell_{long}$.

(a) temperature

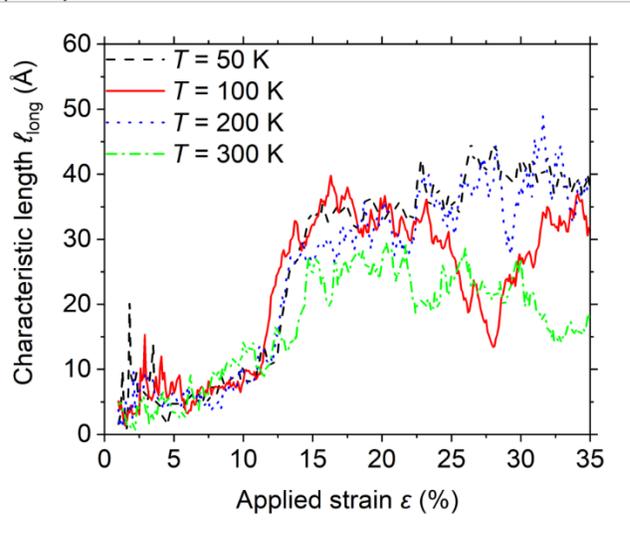

(b) composition

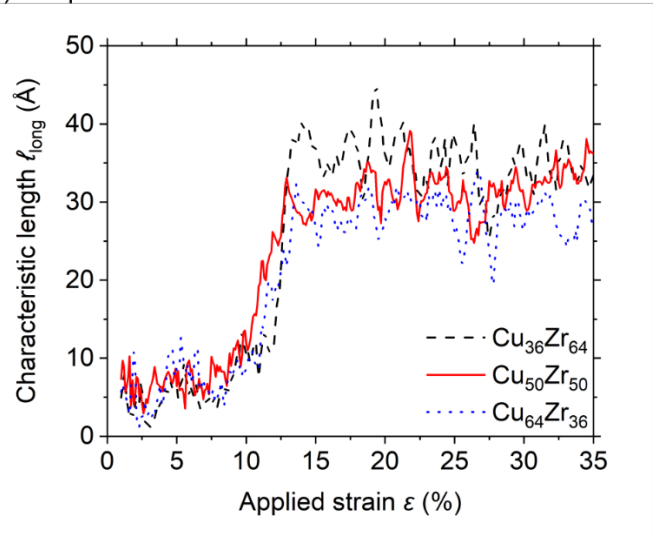

**Figure 8:** Characteristic length $\ell_{long}$ for (a) varying temperature and (b) varying composition. All results are obtained with an offset $\Delta\varepsilon = 1\%$ and $r_{cut} = 3\ r_{NN}$.